\documentclass[prb,twocolumn]{revtex4}
\usepackage{bm}
\usepackage{dcolumn}
\usepackage{graphicx}
\begin{document}

\title{Transient cavities and the excess chemical potentials of
hard-spheroid solutes in dipolar hard sphere solvents}

\author{Philip J. Camp}

\email[E-mail:]{philip.camp@ed.ac.uk}

\affiliation{School of Chemistry, University of Edinburgh, West Mains
Road, Edinburgh EH9 3JJ, United Kingdom}

\date{\today}

\begin{abstract} 
Monte Carlo computer simulations are used to study transient cavities and
the solvation of hard-spheroid solutes in dipolar hard sphere solvents.  
The probability distribution of spheroidal cavities in the solvent is
shown to be well described by a Gaussian function, and the variations of
fit parameters with cavity elongation and solvent properties are analyzed.  
The excess chemical potentials of hard-spheroid solutes with aspect ratios
$x$ in the range $1/5 \leq x \leq 5$, and with volumes between one and
twenty times that of a solvent molecule, are presented. It is shown that
for a given molecular volume and solvent dipole moment (or temperature) a
spherical solute has the lowest excess chemical potential and hence the
highest solubility, while a prolate solute with aspect ratio $x$ should be
more soluble than an oblate solute with aspect ratio $1/x$. For a given
solute molecule, the excess chemical potential increases with increasing
temperature; this same trend is observed in the case of hydrophobic
solvation. To help interpret the simulation results, comparison is made
with a scaled-particle theory that requires prior knowledge of a
solute-solvent interfacial tension and the pure-solvent equation of state,
which parameters are obtained from simulation results for spherical
solutes. The theory shows excellent agreement with simulation results over
the whole range of solute elongations considered.
\end{abstract}

\maketitle

\section{Introduction}
\label{sec:introduction}

The solvation of solutes in simple polar solvents is of importance in
almost all areas of chemistry and biochemistry. The way in which solvent
molecules are ordered (or not) around solute molecules can confer all
manner of interesting and useful phenomena upon the solution, including
catalytic activity, preferential solvation in solute mixtures, microphase
formation, and self-assembly. Some of the most important - and complex -
solvation phenomena occur in aqueous solutions of hydrophobic solutes.
Hydrophobic solvation is characterized by negative energy and entropy
changes accompanying the insertion of hydrophobic solutes in to aqueous
phases. One contribution to the energy change is from the ever-present
attractive dispersion interactions that operate between all species, but
the dominant contribution is from the specific solvent-solvent
interactions facilitated by a reorganization (ordering) of the solvent
molecules around each solute molecule; this latter effect also gives rise
to the negative entropy change. As a result of these changes in energy and
entropy, the solubility of a hydrophobic solute in aqueous solution
decreases with increasing temperature over quite a wide temperature range;  
in the case of methane in water the solubility reaches a minimum at $T
\simeq 350~{\rm K}$. The hydrophobic attraction between hydrophobic
solutes in aqueous solution is an effective interaction arising from the
fact that favorable solvent-solvent interactions are maximized when the
solute molecules are in close proximity with one another, as opposed to
being completely surrounded by solvent. For a recent discussion of these
concepts, and references to the enormous literature in this area, see
Ref.~\onlinecite{Widom:2003/a}.

Considerable effort has been directed toward understanding the nature of
solvation in aqueous solutions, especially as compared to that in simple
non-aqueous solvents. For example, inert gases are less soluble in water
than they are in non-aqueous molecular
liquids.\cite{Pierotti:1963/a,Pierotti:1965/a,Lucas:1976/a} Sophisticated
molecular theories have been proposed specifically for hydrophobic
solvation and attraction.\cite{Pratt:1977/a,Lum:1999/a,Pratt:2002/a} It
was pointed out that the distribution of transient cavities in the pure
solvent could be used to characterize such effects as hydrophobic
solvation.\cite{Lee:1985/a,Lee:1985/b} In the early 1990's, computer
simulations were used to characterize the transient cavities in a range of
molecular liquids.\cite{Pohorille:1990/a,Pratt:1992/a} This was achieved
by calculating the probability of finding a spherical solute-sized cavity
in the solvent. It was found that although water has a larger free volume
-- or alternatively, a smaller effective packing fraction -- than do
non-aqueous solvents, that free volume is distributed throughout smaller
cavities. Moreover, non-aqueous solvents exhibit a greater ability to
redistribute free volume in order to accommodate large solute species.
Such considerations are of fundamental importance in understanding the
hydrophobic interaction and all of its manifestations in biochemistry,
nanoscale systems, and materials chemistry.

This very brief discussion highlights the value of understanding the
molecular-scale cavity structure in the pure solvents before one proceeds
to consider specific solvent-solute systems. In
Ref.~\onlinecite{Lee:1985/b} it was suggested that such investigations
could be extended to non-spherical cavities and solutes. Simulation
studies of this type have been carried out for various hard-core
solute/solvent systems
\cite{deSouza:1994/a,deSouza:1994/b,Stamatopoulou:1995/a,%
Stamatopoulou:1997/a,Stamatopoulou:1998/a,Omelyan:2001/a} but we are not
aware of any systematic comparison between rod-like and disk-like solutes.
The current work is therefore concerned with transient non-spherical
cavities and the solvation of non-spherical solutes in model polar
solvents. For simplicity, we consider hard-spheroid solutes (or uniaxial
hard ellipsoids) in a solvent made up of dipolar hard spheres (DHSs). The
DHS system is the simplest model of a polar liquid, and is of considerable
intrinsic interest owing to its complex phase behavior.
\cite{Weis:1993/a,Weis:1993/b,Levesque:1994/a,PJC:2000/a,PJC:2000/c,
Teixeira:2000/a,Tlusty:2000/a,Zilman:2003/a,Huke:2004/a} In particular,
the low-temperature properties of the DHS fluid are dominated by the
association of particles in `nose-to-tail' conformations giving rise to
chains and rings at low densities, and extended networks at intermediate
densities.\cite{Weis:1993/a,Levesque:1994/a,PJC:2000/c} Therefore, the DHS
system is an interesting example of a fluid that should exhibit large
transient cavities as compared to non-polar hard-sphere fluids at the same
density. We use Monte Carlo (MC) computer simulations to study the size
distributions of prolate (rod-like) and oblate (disk-like) spheroidal
cavities in DHS fluids, and relate these distributions to the excess
chemical potentials of hard-spheroid solutes. This allows an examination
of the role of solute shape, as well as solute volume, on the solubilities
of non-polar solutes in polar fluids. Following earlier simulation studies
of this type, we compare our calculations with a scaled-particle theory
(SPT) which describes the reversible work for insertion and subsequent
growth of a single solute molecule in to a solvent.

This article is organized as follows. In Section \ref{sec:modelandmethods}
we describe the solute and solvent models, the SPT as applied to the model
system, and the computer simulation methods employed in this work. Results
are presented in Section \ref{sec:results}, and Section
\ref{sec:conclusions} concludes the paper.

\section{Model and methods}
\label{sec:modelandmethods}

\subsection{Model}
\label{sec:model}

The DHS solvent consists of hard spheres with diameter $\sigma$, each
carrying a central dipole moment ${\bm\mu}$. For two spheres with
separation vector ${\bm r}$, the pair potential energy is
\begin{equation}
u({\bm r},{\bm\mu}_{1},{\bm\mu}_{2}) = 
\left\{
 \begin{array}{ll}
  \infty 
  & r < \sigma \\
  \frac{({\bm\mu}_{1}\cdot{\bm\mu}_{2})}
       {4\pi\epsilon_{0}r^{3}}-
  \frac{3({\bm\mu}_{1}\cdot{\bm r})({\bm\mu}_{2}\cdot{\bm r})}
       {4\pi\epsilon_{0}r^{5}} 
  & r \geq \sigma
 \end{array}
\right.
\label{eqn:u}
\end{equation}
where $r=\left| {\bm r} \right|$ and $\epsilon_{0}$ is the vacuum
dielectric permitivity. Thermodynamic parameters of the DHS system are
expressed in dimensionless form as follows: the reduced number density
$\rho^{*}=N\sigma^{3}/V$, where $N$ is the number of particles in a volume
$V$; the reduced dipole moment
$\mu^{*}=\sqrt{\mu^{2}/4\pi\epsilon_{0}k_{B}T\sigma^{3}}$, where $k_{B}$
is Boltzmann's constant, and $T$ is the temperature; the reduced
temperature $T^{*}=(1/\mu^{*})^{2}$. Throughout the rest of the paper, the
DHSs will be referred to as component `1'.

We will compute the probability of finding a spheroidal cavity in the DHS
fluid centered at a randomly selected point in the system, and hence
obtain the excess chemical potentials of hard-spheroid solutes. The solute
will be referred to as component `2' of the resulting solution. With the
spheroid radial semi-axes denoted by $a$, and the polar semi-axis by $c$,
the elongation of the spheroid is given by $x=c/a$. The fundamental
measures of the spheroid will be required in Section \ref{sec:spt}, and
are therefore recorded in Table \ref{tab:spheroid} for reference.
\begin{table*}
\begin{center}
\caption{\label{tab:spheroid} Mean radius of curvature, surface area,
and volume for prolate and oblate spheroids with radial semi-axis $a$,
polar semi-axis $c$, and elongation $x=c/a$.}
\begin{tabular}{llcc}\hline\hline\\
 & & Prolate ($x>1$) & Oblate ($x<1$) \\ \hline
Eccentricity & $\epsilon$ &
$\sqrt{1-x^{-2}}$ &
$\sqrt{1-x^{2}}$ \\
Mean radius of curvature & $R_{2}$ &
$\frac{c}{2}
\left[
 1 +
 \frac{1-\epsilon^{2}}{2\epsilon}
 \ln{\left(\frac{1+\epsilon}{1-\epsilon}\right)}
\right]$ &
$\frac{c}{2}
\left[
 1 +
 \frac{\arcsin{\epsilon}}{\epsilon\sqrt{1-\epsilon^{2}}}
\right]$ \\
Surface area & $S_{2}$ &
$2\pi a^{2}
\left[
 1 +
 \frac{\arcsin{\epsilon}}{\epsilon\sqrt{1-\epsilon^{2}}}
\right]$ &
$2\pi a^{2}
\left[
 1 +
 \frac{1-\epsilon^{2}}{2\epsilon}
 \ln{\left(\frac{1+\epsilon}{1-\epsilon}\right)}
\right]$ \\
Volume & $V_{2}$ &
$\frac{4}{3}\pi a^{2}c$ &
$\frac{4}{3}\pi a^{2}c$ \\
\hline\hline
\end{tabular}
\end{center}
\end{table*}

\subsection{Thermodynamics}
\label{sec:thermodynamics}

Of central interest in this paper is the excess chemical potential,
$\Delta\mu_{2}$, of hard-spheroid solutes in DHS solvents at low
concentrations where solute-solute interactions can be ignored.  
$\Delta\mu_{2}$ can be related to the solubility of the solute in a
variety of situations, but to provide a concrete example consider the
transfer of a solute molecule from an ideal gas to an initially pure
solvent (or an extremely dilute solution) such that the volumes of the gas
and the solvent separately remain constant. Thermal equilibrium between
the gas and the solution is assumed. The chemical potential of the solute
in the ideal gas is equal to
\begin{equation}
\mu_{2}^{\rm vap} = k_{B}T\ln{\rho_{2}^{\rm vap}{\cal V}}
\label{eqn:mu2vap}
\end{equation}
where $\rho_{2}^{\rm vap}$ is the number density of solutes in the gas
phase, and ${\cal V}$ is the de Broglie thermal volume of the solute. The
chemical potential of solute molecules in solution (in the absence of
solute-solute interactions) is written
\begin{equation}
\mu_{2} = k_{B}T\ln{\rho_{2}{\cal V}}
        + \Delta\mu_{2}
\label{eqn:mu2soln}
\end{equation}
where $\rho_{2}$ is the number density of solutes in the solution. When
the solute is at chemical equilibrium the chemical potentials of the
solute in both phases are equal, and from the mass-action law the
associated partition coefficient can be defined as
\begin{equation}
K \equiv \frac{\rho_{2}}{\rho_{2}^{\rm vap}}
  = \exp{(-\beta\Delta\mu_{2})}.
\label{eqn:K}
\end{equation}
where $\beta=1/k_{B}T$. The temperature dependence of the partition
coefficient yields information on the energetic and entropic contributions
to $\beta\Delta\mu_{2}$. For transfer under conditions of constant volume
the relevant expressions are
\begin{equation}
\Delta u_{2} = k_{B}T^{2}\frac{\partial\ln{K}}{\partial T}
\label{eqn:deltau2}
\end{equation}
for the energy, and
\begin{equation}
\Delta s_{2} = k_{B}\frac{\partial(T\ln{K})}{\partial T}
\label{eqn:deltas2}
\end{equation}
for the entropy, where all differentiations are carried out with gas
volume, solvent volume, and number of solvent molecules held constant.
Widom {\it et al.} have provided full details of analogous expressions for
transfer under conditions of constant pressure, and the physical
situations to which Eqs.~(\ref{eqn:deltau2}) and (\ref{eqn:deltas2}) are
applicable. \cite{Koga:2002/a,Widom:2003/a}

\subsection{Scaled particle theory}
\label{sec:spt}

We note that there are several theories of hard solute/solvent systems
that rely on representing non-spherical components by spherical particles
with effective
radii;\cite{deSouza:1994/a,deSouza:1994/b,Stamatopoulou:1995/a,%
Stamatopoulou:1997/a,Stamatopoulou:1998/a,Omelyan:2001/a} the resulting
effective hard-sphere mixture is then described by accurate semi-empirical
formulae, such as the Boublik-Monsoori-Carnahan-Starling-Leland equation
of state.\cite{Boublik:1970/a,Mansoori:1971/a} In this work, we will show
that the variations of $\Delta\mu_{2}$ with solute size and shape can be
described most intuitively within a simple version of
SPT.\cite{Reiss:1959/a} The excess chemical potential of a single hard
spheroid with radial semi-axis $a$ and polar semi-axis $c$ is equal to the
reversible work required to insert a point particle in to the solvent, and
then to grow the point particle to full size. Consider, then, a scaled
spheroid with radial and polar semi-axes equal to $\lambda a$ and $\lambda
c$, respectively, with $\lambda=0$ corresponding to the point particle,
and $\lambda=1$ to the full-sized particle. For hard-particle solutes and
solvents, the Widom formula relates the excess chemical potential for the
solute at reciprocal temperature $\beta=1/k_{B}T$ to the probability, $P$,
of inserting a solute at random in to the solution without incurring any
overlaps:\cite{Widom:1963/a}
\begin{equation}
\beta\Delta\mu_{2} = -\ln{P}.
\label{eqn:mu2widom}
\end{equation}
In the spirit of SPT, we will interpolate between results that are correct
in the limits $\lambda=0$ and $\lambda=\infty$ to obtain a result for
$\lambda=1$. When $\lambda$ is small, the chemical potential of the scaled
spheroid is related to the free volume in the solvent, i.e.
\begin{equation}
\beta\Delta\mu_{2} \approx 
-\ln{\left[\frac{V-NV_{\rm ex}(\lambda)}{V}\right]}
\label{eqn:mu2low}
\end{equation}
where $V_{\rm ex}(\lambda)$ is the excluded volume of the scaled spheroid
and a single solvent molecule. Obviously, this is only correct when the
solute is sufficiently small that it can only be in the vicinity of one
solvent molecule at any given time. The sphere-scaled spheroid excluded
volume is given by the following expression due to
Kihara:\cite{Kihara:1963/a}
\begin{equation}
V_{\rm ex}(\lambda) = V_{1}
                    + \lambda S_{1}R_{2}
                    + \lambda^{2} R_{1}S_{2}
                    + \lambda^{3} V_{2}
\label{eqn:vex}
\end{equation}
The fundamental measures for spheroids are given in Table
\ref{tab:spheroid}; for DHSs the measures are $R_{1}=\sigma/2$,
$S_{1}=\pi\sigma^{2}$, $V_{1}=\pi\sigma^{3}/6$. In the other extreme,
$\lambda\rightarrow\infty$, the reversible work for inserting the solute
particle is dominated by the work done against the pressure, $p$, of the
solvent, and the (normally positive) contribution to the free energy
arising from the interface between the solute molecule and the solvent. In
this case the excess chemical potential is
\begin{equation}
\beta\Delta\mu_{2} \approx \frac{\lambda^{2}\gamma S_{2}}{k_{B}T} +
\frac{\lambda^{3}pV_{2}}{k_{B}T}
\label{eqn:mu2high}
\end{equation}
where $\gamma$ is the solute-solvent interfacial tension. We now
approximate the excess chemical potential of a full-sized solute molecule
with a cubic function of $\lambda$: the terms up to order $\lambda$ are
obtained by Taylor expansion of the combination of (\ref{eqn:mu2low}) and
(\ref{eqn:vex}); and the terms in $\lambda^{2}$ and $\lambda^{3}$ are
taken from (\ref{eqn:mu2high}). Finally, we set $\lambda=1$ (corresponding
to the full-sized solute molecule) which yields
\begin{equation}
\beta\Delta\mu_{2} = -\ln{(1-\eta)}
                   + \frac{6\eta}{1-\eta}\left(\frac{R_{2}}{\sigma}\right)
                   + \frac{\gamma S_{2}}{k_{B}T}
                   + \frac{pV_{2}}{k_{B}T}
\label{eqn:mu2spt}
\end{equation}
where $\eta=\rho V_{1}$ is the packing fraction of the pure solvent. The
first term in (\ref{eqn:mu2spt}) arises from the fact that the probability
of inserting a point particle in to the solvent without overlap is equal
to $(1-\eta)$. (The surface term proportional to $S_{2}$ can also contain
an additional curvature contribution\cite{Rowlinson:2003/a} but it will be
explained below that this is unnecessary in the current application.) The
pressure and solute-solvent interfacial tension will be determined from
computer simulations which we describe in the next section.

\subsection{Computer simulations}
\label{sec:simulations}

Canonical ($NVT$) MC simulations of $N=500$ DHSs were carried out in a
cubic simulation cell with periodic boundary conditions
applied.\cite{Allen:1987/a} The long-range dipolar interaction was handled
using the Ewald summation with conducting boundary conditions. One MC
cycle consisted of one trial translation and one trial rotation per
molecule, on average. Run lengths after equilibration consisted of ${\cal
O}(10^{5})$ MC cycles. The respective maximum displacements for
translational and rotational moves were adjusted to give acceptance rates
of 20\%.

To characterize the cavity structure in the DHS fluid, a test particle was
inserted in to the simulation cell with randomly selected position and
orientation. For a given elongation, the maximum possible spheroid radial
semi-axis, $a_{m}$, that would not result in any particle overlaps was
determined. This procedure was repeated 500 times every 20 MC cycles, and
a histogram of $a_{m}$ was accumulated to yield a cavity function
$Q(a_{m})$, the integral of which is
\begin{equation}
\int_{0}^{\infty} Q(a_{m}) da_{m} = 1 - \eta.
\label{eqn:Q}
\end{equation}
In addition, the excess chemical potentials
$\beta\Delta\mu_{2}=-\ln{P(a)}$ were determined using (\ref{eqn:mu2widom})
and the fact that for a given test-particle insertion, all spheroids with
the same elongation and $a\leq a_{m}$ would be accommodated in the solvent
without overlap. $P(a)$ and $Q(a_{m})$ are related because the probability
of successfully inserting a test particle with radial semi-axis $a$ is
equal to the likelihood of there being a cavity that can accommodate a
particle with radial semi-axis of at least that size,
\cite{Pohorille:1990/a} i.e.
\begin{equation}
P(a) 
= \int_{a}^{\infty} Q(a_{m}) da_{m} 
= 1 - \eta - \int_{0}^{a} Q(a_{m}) da_{m}.
\label{eqn:PQ}
\end{equation}
As will be explained in Section \ref{sec:parametrization}, MC simulation
results were used to parametrize the SPT expression (\ref{eqn:mu2spt}).
The pressure in the DHS fluid was determined using the virial
equation,\cite{Hansen:1986/a}
\begin{equation}
\frac{p}{\rho k_{B}T} = 1
                      + \frac{2}{3}\pi\rho\sigma^{3}g(\sigma)
                      + \frac{U}{Nk_{B}T}
\label{eqn:pVnkT}
\end{equation}
where $g(\sigma)$ is the radial distribution function, $g(r)$, at contact,
and $U$ is the dipolar configurational energy. The second term in
(\ref{eqn:pVnkT}) represents the hard-core contribution to the equation of
state; $g(\sigma)$ was estimated by extrapolating $g(r)$ to contact using
a $[2,2]$ Pad{\'e} approximant.

\section{Results}
\label{sec:results}

The DHS fluid was studied with reduced dipole moments $\mu^{*}=0$ (hard
spheres), $\mu^{*}=1$, and $\mu^{*}=2$, corresponding to reduced
temperatures of $T^{*}=\infty$, $T^{*}=1$, and $T^{*}=0.25$, respectively.  
The number densities considered were $\rho^{*}=0.2$, $\rho^{*}=0.5$, and
$\rho^{*}=0.8$. (For water under ambient conditions, $\mu^{*} \sim 2$ and
$\rho^{*} \sim 0.9$; rough estimates of the parameters for liquid ammonia
are $\mu^{*} \sim 1$ and $\rho^{*} \sim 0.8$.) Hard-spheroid solute
particles were considered with elongations in the range $1/5 \leq x \leq
5$, and with a range of molecular volumes up to $20V_{1}$, depending on
the density; at higher densities the probability of inserting a large
solute particle in to the solvent without overlap becomes too small to
enable an accurate evaluation of (\ref{eqn:mu2widom}).

\subsection{Cavity distributions and molecular structure}
\label{sec:cavity}

\begin{figure}[!tbp]
\centering   
\includegraphics[angle=270,scale=0.30]{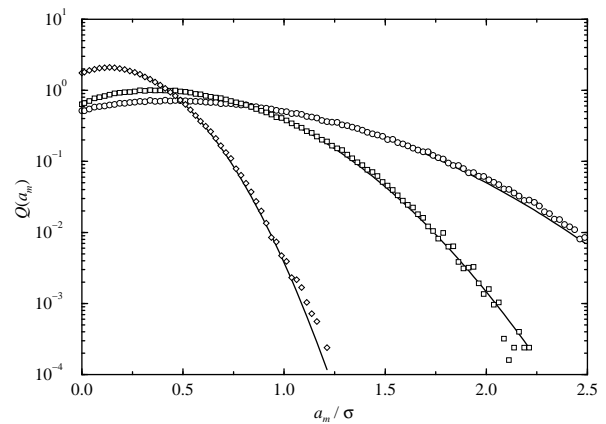}
\caption{\label{fig:gaussians} Linear-log plot of the cavity distribution
functions, $Q(a_{m})$, for spheroidal cavities of radial semi-axis $a$ and
elongation $x$ in a DHS solvent with $\rho^{*}=0.2$ and $\mu^{*}=2$:
$x=1/5$ (circles); $x=1$ (squares); $x=5$ (diamonds). The solid lines are
fits to the Gaussian distribution.}
\end{figure}

\begin{figure}[!tbp]
\centering   
\includegraphics[angle=270,scale=0.30]{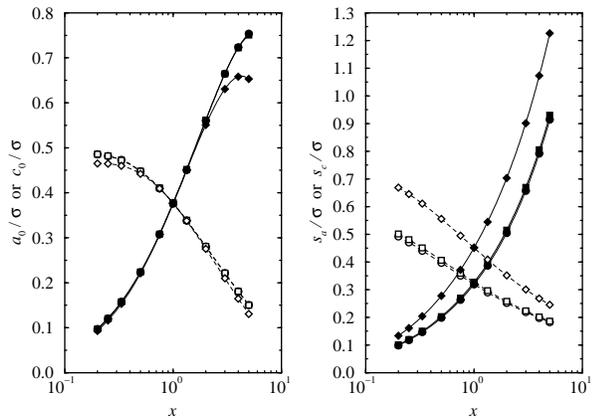}
\caption{\label{fig:r0.2avs} Statistics for the radial semi-axis (open
symbols) and polar semi-axis (filled symbols) of spheroidal cavities in
DHS solvents at density $\rho^{*}=0.2$ and dipole moments $\mu^{*}=0$
(circles), $\mu^{*}=1$ (squares), and $\mu^{*}=2$ (diamonds): (left) most
probable values, $a_{0}$ and $c_{0}$; (right) width parameters, $s_{a}$
and $s_{c}$. Error bars are smaller than the symbols.}
\end{figure}

\begin{figure}[!tbp]
\centering   
\includegraphics[angle=270,scale=0.30]{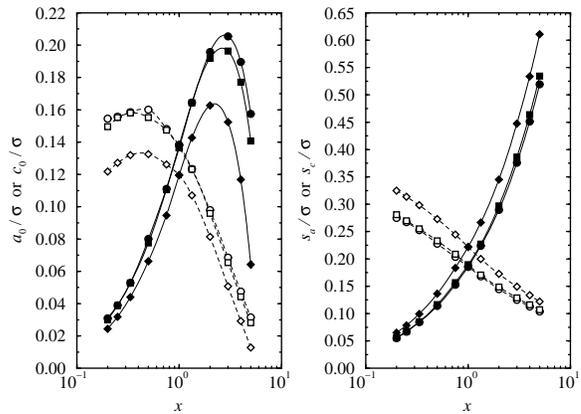}
\caption{\label{fig:r0.5avs} Statistics for the radial semi-axis (open
symbols) and polar semi-axis (filled symbols) of spheroidal cavities in
DHS solvents at density $\rho^{*}=0.5$ and dipole moments $\mu^{*}=0$
(circles), $\mu^{*}=1$ (squares), and $\mu^{*}=2$ (diamonds): (left) most
probable values, $a_{0}$ and $c_{0}$; (right) width parameters, $s_{a}$
and $s_{c}$. Error bars are smaller than the symbols.}
\end{figure}

\begin{figure}[!tbp]
\centering
\includegraphics[angle=270,scale=0.30]{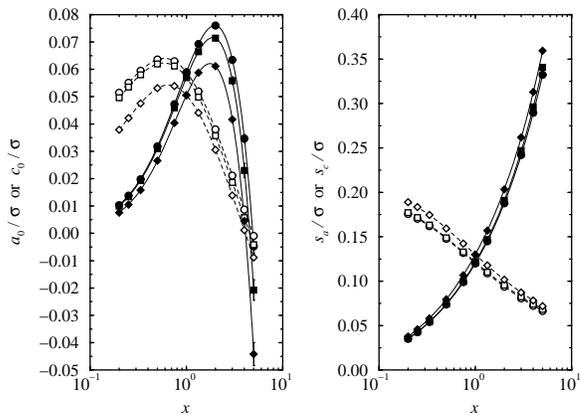}
\caption{\label{fig:r0.8avs} Statistics for the radial semi-axis (open
symbols) and polar semi-axis (filled symbols) of spheroidal cavities in
DHS solvents at density $\rho^{*}=0.8$ and dipole moments $\mu^{*}=0$
(circles), $\mu^{*}=1$ (squares), and $\mu^{*}=2$ (diamonds): (left) most
probable values, $a_{0}$ and $c_{0}$; (right) width parameters, $s_{a}$
and $s_{c}$. Error bars are smaller than the symbols.}
\end{figure}

In Fig.~\ref{fig:gaussians} we show the cavity distributions, $Q(a_{m})$,
for spheroidal cavities in a DHS solvent with $\rho^{*}=0.2$ and
$\mu^{*}=2$. Results are presented for spheroids with elongations of
$x=1/5$ (oblate), $x=1$ (spherical), and $x=5$ (prolate). The
distributions are well described by the Gaussian function
\begin{equation}
Q(a_{m}) \propto 
\exp{\left[-\frac{1}{2}\left(\frac{a_{m}-a_{0}}{s_{a}}\right)^{2}\right]}
\label{eqn:Qfit}
\end{equation}
where $a_{0}$ is the most probable radial semi-axis, and $s_{a}$ is the
corresponding width parameter. Distributions of the maximum polar
semi-axis are defined analogously. The Gaussian function produced
excellent fits to all of the DHS state-points and cavity elongations
studied; results for $\rho^{*}=0.2$ and $\mu^{*}=2$ are included in
Fig.~\ref{fig:gaussians}. In Figs.~\ref{fig:r0.2avs}, \ref{fig:r0.5avs},
and \ref{fig:r0.8avs} we show the fitting parameters for DHS solvents at
densities of $\rho^{*}=0.2$, $\rho^{*}=0.5$, and $\rho^{*}=0.8$,
respectively, as functions of the spheroid elongation. These figures also
include the corresponding parameters for the polar semi-axis, $c$.

The results for $\rho^{*}=0.2$ (Fig.~\ref{fig:r0.2avs}) show that with all
dipole moments, the most probable radial (polar) semi-axis decreases
(increases) with increasing elongation. In addition, the most probable
polar semi-axis for prolate cavities with elongation $x>1$ is greater than
the most probable radial semi-axis for oblate cavities with elongation
$1/x<1$; this is just due to fact that the polar semi-axis of a long, thin
spheroid is longer than the radial semi-axis of a short, fat spheroid with
the same volume. For a given cavity elongation, $a_{0}$ and $c_{0}$ are
almost independent of the solvent polarity (except at the highest
elongations considered where small deviations are apparent). The width
parameters in the Gaussian functions vary in a similar way to the
corresponding most probable values, although they appear to be more
sensitive to the solvent dipole moment. Results for $\mu^{*}=0$ and
$\mu^{*}=1$ are almost identical, but those for $\mu^{*}=2$ are
significantly larger for both semi-axes and all elongations. At low dipole
moments (or high temperatures) the distribution of cavity sizes is
narrower since there is only one characteristic lengthscale in the
solvent. At high dipole moments there is strong association of the solvent
particles which gives rise to two structural lengthscales: the average
distance between nearest-neighbor solvent particles, and the
characteristic lengthscale of the cluster network.\cite{PJC:2000/c} This
gives rise to a heterogeneous cavity structure in which small cavities are
located in the vicinity of clustered solvent particles, and larger voids
are formed in the free space between clusters. Hence, a broader
distribution of cavity sizes is found at higher dipole moments. The width
parameter is clearly a sensitive probe of the solvent structure.

Results for $\rho^{*}=0.5$ (Fig.~\ref{fig:r0.5avs}) and $\rho^{*}=0.8$
(Fig.~\ref{fig:r0.8avs}) show similar trends, except that the most
probable dimensions show local maxima in the range of elongations
considered. For a given solvent dipole moment, these maximal values of
$a_{0}$ and $c_{0}$ occur for less anisotropic shapes as the density is
increased, which suggests that there are fewer long `holes' in the fluid
structure at high densities. The most probable cavity dimensions and
associated width parameters decrease with increasing density. This is
because as the density is increased, the free volume in the vicinity of
the clustered solvent molecules represents a greater proportion of the
total free volume; since small cavities are located in the region of the
solvent molecules, and large cavities are accommodated in the voids of the
clustered network structure, the average cavity dimensions and associated
`standard deviations' will be reduced. $a_{0}$ and $c_{0}$ decrease with
increasing solvent dipole moment, whereas $s_{a}$ and $s_{c}$ increase,
once again due to the promotion of solvent-particle association. The
development of a heterogeneous fluid structure gives rise to a greater
spread of cavity sizes, whereas the shifts in $a_{0}$ and $c_{0}$ must be
due to the decreasing nearest-neighbor separation between solvent
particles.

\begin{figure}[!tbp]
\centering   
\includegraphics[scale=0.40]{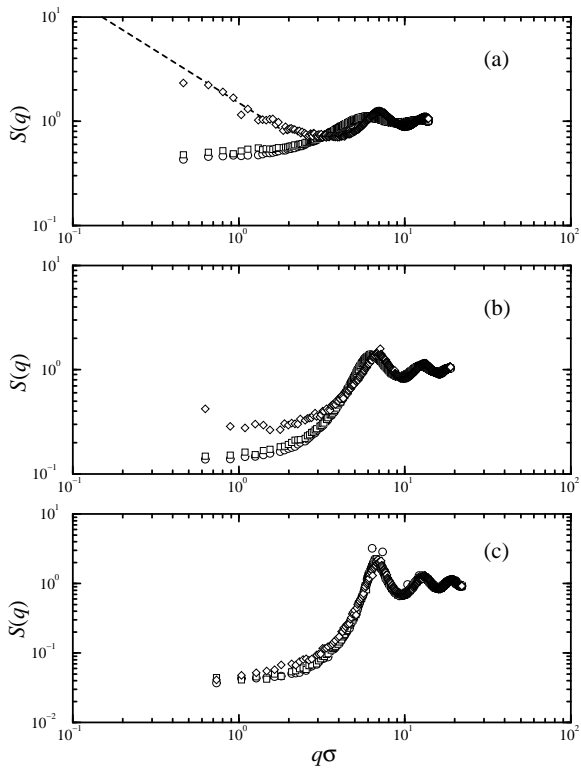}
\caption{\label{fig:sq} Log-log plots of the static structure factors,
$S(q)$, for DHS solvents with dipole moments $\mu^{*}=0$ (circles),
$\mu^{*}=1$ (squares), and $\mu^{*}=2$ (diamonds): (a) $\rho^{*}=0.2$; (b)
$\rho^{*}=0.5$; (c) $\rho^{*}=0.8$. The dashed line in (a) follows the
power law $S(q) \propto q^{-1}$.}
\end{figure}

The influences of solvent dipole moment on the fit parameters are seen to
decrease in magnitude with increasing density. This is due to the fact
that at low densities strongly interacting DHSs associate to form networks
that possess heterogeneous cavity structures, whereas at high densities
the fluid structure is not qualitatively different from that of pure hard
spheres. This statement is backed up by Fig.~\ref{fig:sq}, which shows the
static structure factor, $S(q)$, in DHS solvents at each density and
dipole moment considered in this work. There are many alternative choices
of structural probe, but $S(q)$ does have the merit of observing certain
characteristic scaling laws in the presence of strong particle
association.\cite{PJC:2000/c} $S(q)$ was calculated in the simulations
directly using the relation
\begin{equation}
S({\bm q}) =
\frac{1}{N} 
\left\langle
 \left[ \sum_{i=1}^{N}\cos{({\bm q}\cdot{\bm r}_{i})} \right]^{2}
+\left[ \sum_{i=1}^{N}\sin{({\bm q}\cdot{\bm r}_{i})} \right]^{2}
\right\rangle
\label{eqn:sq}
\end{equation}
where ${\bm q}$ is a reciprocal lattice vector of the simulation cell, and
contributions with equal $q=\left|{\bm q}\right|$ were averaged. At a
density of $\rho^{*}=0.2$ the structure of the system with $\mu^{*}=2$ is
clearly different from the structures of those systems with $\mu^{*}=1$
and $\mu^{*}=0$. In particular, the strongly polar system exhibits
power-law behavior at low $q$; as discussed fully in
Ref.~\onlinecite{PJC:2000/c}, chain-like correlations give rise to the
scaling $S(q)\sim q^{-1}$. At the higher densities, the structural
features of the fluid are far less sensitive to the dipole moment.

\subsection{Parametrizing the SPT}
\label{sec:parametrization}

\begin{figure}[!tbp]
\centering   
\includegraphics[angle=270,scale=0.30]{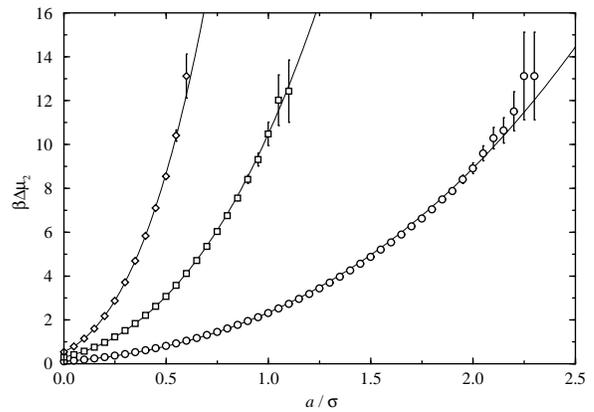}
\caption{\label{fig:spt} Excess chemical potentials of hard-sphere solutes
with radius $a$ in DHS solvents with $\mu^{*}=2$ and at various densities:
$\rho^{*}=0.2$ (circles); $\rho^{*}=0.5$ (squares); $\rho^{*}=0.8$
(diamonds). The symbols are from MC simulations and the curves are best
fits using (\ref{eqn:mu2spt})}.
\end{figure}

To test the SPT expression in (\ref{eqn:mu2spt}) we need to determine the
values of the solute-solvent interfacial tension and the solvent equation
of state. The latter was obtained as described in Section
\ref{sec:simulations}, and the results are included in Table
\ref{tab:thermo}. The interfacial tension was obtained by fitting
(\ref{eqn:mu2spt}) to simulation results for spherical solutes only, with
the intention of then applying the parametrized equation to non-spherical
solutes. As examples of the procedure, in Fig.~\ref{fig:spt} we show the
excess chemical potentials of hard-sphere solutes as functions of the
radius, $a$, in DHS systems with $\mu^{*}=2$. It is emphasized that the
only fitting parameter in (\ref{eqn:mu2spt}) is the reduced interfacial
tension $\gamma\sigma^{2}/k_{B}T$.\footnote{Note that in the particular
case of the DHS solvent there is some uncertainty as to the location and
nature of any vapor-liquid phase transitions and so $\gamma$ cannot be
approximated by a known surface tension. In any case, current
estimates\cite{PJC:2000/a} of the `critical temperature' correspond to
reduced dipole moments of $\mu^{*}\simeq 2.5$ and so the solvents used
here can be considered supercritical.} The fits included in
Fig.~\ref{fig:spt} are seen to be quite good, and the resulting fit
parameters are reported in Table \ref{tab:thermo}. In earlier
works,\cite{Pohorille:1990/a,Pratt:1992/a} it has sometimes proven
necessary to modify the interfacial term with a curvature correction, i.e.
the third term on the right-hand side of (\ref{eqn:mu2spt}) is replaced by
a contribution like\cite{Rowlinson:2003/a}
\begin{equation} 
\frac{\gamma S_{2}}{k_{B}T}
\left( 1 - \frac{4\delta}{R_{2}} \right) 
\label{eqn:curvature} 
\end{equation} 
where $\delta$ is also to be parametrized against simulation results. This
modification was fully tested against simulation results, but in some
cases it led to negative interfacial tensions, and in every case gave rise
to significant uncertainties in the fit parameters with no visible
improvement over (\ref{eqn:mu2spt}).
\begin{table*}
\begin{center}
\caption{\label{tab:thermo} Thermodynamic properties of the DHS fluid: the
reduced density $\rho^{*}$; the reduced dipole moment $\mu^{*}$; the
reduced temperature $T^{*}$; the radial distribution function at contact
$g(\sigma)$; the dipolar energy, $U$, per molecule in units of $k_{B}T$;
the reduced pressure $p\sigma^{3}/k_{B}T$; the reduced solute-solvent
interfacial tension $\gamma\sigma^{2}/k_{B}T$.}
\begin{tabular}{ddddddd}\hline\hline
\multicolumn{1}{r}{$\rho^{*}$} &
\multicolumn{1}{r}{$\mu^{*}$} &
\multicolumn{1}{r}{$T^{*}$} &
\multicolumn{1}{r}{$g(\sigma)$} &
\multicolumn{1}{r}{$U/Nk_{B}T$} &
\multicolumn{1}{r}{$p\sigma^{3}/k_{B}T$} &
\multicolumn{1}{r}{$\gamma\sigma^{2}/k_{B}T$} \\ \hline
0.2 & 0 & \multicolumn{1}{c}{$\infty$}
                   & 1.3220 &  0      & 0.3107 & 0.13103(2) \\
0.2 & 1 & 1        & 1.6492 & -0.2847 & 0.2812 & 0.13085(2) \\
0.2 & 2 & 0.25     & 9.6583 & -4.6588 & 0.0774 & 0.09579(7) \\
0.5 & 0 & \multicolumn{1}{c}{$\infty$}
                   & 2.1558 &  0      & 1.6288 & 0.5128(1)  \\
0.5 & 1 & 1        & 2.4148 & -0.6625 & 1.4332 & 0.5089(1)  \\
0.5 & 2 & 0.25     & 5.6678 & -5.7950 & 0.5702 & 0.4516(1)  \\
0.8 & 0 & \multicolumn{1}{c}{$\infty$}
                   & 4.0094 &  0      & 6.1742 & 1.3653(3)  \\
0.8 & 1 & 1        & 4.2377 & -1.0056 & 5.6758 & 1.3623(6)  \\
0.8 & 2 & 0.25     & 6.1733 & -6.8986 & 3.5558 & 1.3153(4)  \\
\hline\hline
\end{tabular}
\end{center}
\end{table*}

\subsection{Excess chemical potentials of spheroid solutes}
\label{sec:mu2res}

\begin{figure}[!tbp]
\centering   
\includegraphics[angle=270,scale=0.30]{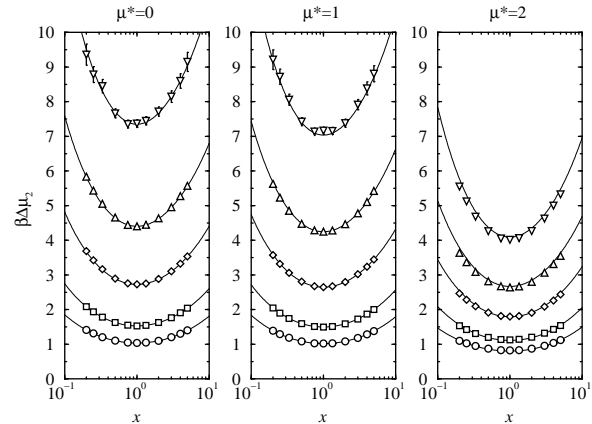}
\caption{\label{fig:r0.2vol} Excess chemical potentials of hard-spheroid
solutes with elongation $x$ in DHS solvents with $\rho^{*}=0.2$ and
various dipole moments: (left) $\mu^{*}=0$; (middle) $\mu^{*}=1$; (right)  
$\mu^{*}=2$. In each case the symbols correspond to solute molecular
volumes of $V_{2}=V_{1}$ (circles), $V_{2}=2V_{1}$ (squares),
$V_{2}=5V_{1}$ (diamonds), $V_{2}=10V_{1}$ (up triangles), and
$V_{2}=20V_{1}$ (down triangles). The symbols are from MC simulations and
the curves are the predictions of (\ref{eqn:mu2spt}).}
\end{figure}

\begin{figure}[!tbp]
\centering   
\includegraphics[angle=270,scale=0.30]{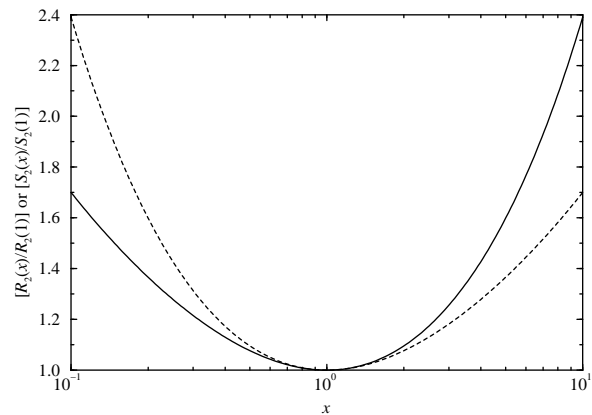}
\caption{\label{fig:measures} Fundamental measures of spheroids compared
to the corresponding measures of spheres with equal volume, as functions
of the elongation $x$: the mean radius of curvature, $R_{2}$ (solid line);
the surface area, $S_{2}$ (dashed line).}
\end{figure}

With (\ref{eqn:mu2spt}) parametrized using simulation results for
spherical solutes, we now turn to an examination of the excess chemical
potentials for non-spherical solutes. To emphasize the effects of
molecular shape we compare results for hard-spheroid solutes with the same
volume, $V_{2}$, but with different elongations (and hence different
$R_{2}$ and $S_{2}$). In Fig.~\ref{fig:r0.2vol} we present simulation
results for DHS solvents at $\rho^{*}=0.2$, and solutes with molecular
volumes in the range $V_{1} \leq V_{2} \leq 20V_{1}$. For a given
molecular volume the spherical solute has the lowest value of
$\beta\Delta\mu_{2}$, and hence from (\ref{eqn:K}) the highest solubility.
This reflects the well-known fact that the ratio of surface area to volume
is smallest for spherical molecules, and hence the unfavorable interfacial
tension term in (\ref{eqn:mu2spt}) is minimized with this geometry.
$\beta\Delta\mu_{2}$ increases sharply as the solute geometry deviates
from the sphere, and there is an approximate correspondence between the
excess chemical potentials of oblate solutes with elongation $1/x<1$ and
prolate solutes with elongation $x>1$, although at more extreme solute
volumes the values for prolate solutes are clearly lower than those for
the corresponding oblate solutes. These variations are well described by
the SPT prediction (\ref{eqn:mu2spt}) even though it was parametrized
against spherical solutes only. Hence we conclude that the variations in
$\beta\Delta\mu_{2}$ are mainly due to a combination of differences in the
mean radius of curvature $R_{2}$, and the surface area $S_{2}$. This is
illustrated in Fig.~\ref{fig:measures} which shows $R_{2}$ and $S_{2}$ as
compared to the corresponding measures of spherical particles of the same
volume. $S_{2}/R_{2}$ increases with increasing particle volume, and so
eventually the variation of $\beta\Delta\mu_{2}$ with $x$ will mirror that
of $S_{2}$ shown in Fig.~\ref{fig:measures}.

\begin{figure}[!tbp]
\centering   
\includegraphics[angle=270,scale=0.30]{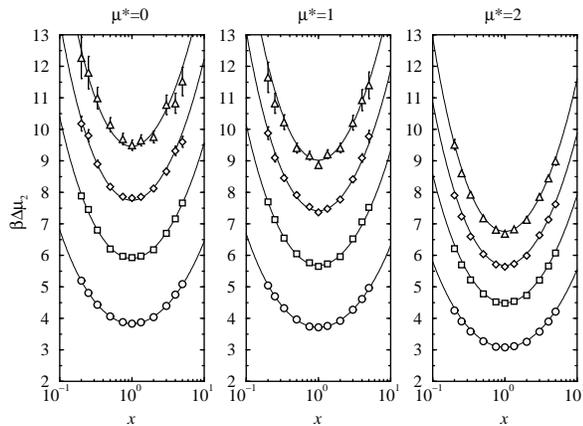}
\caption{\label{fig:r0.5vol} Excess chemical potentials of hard-spheroid
solutes with elongation $x$ in DHS solvents with $\rho^{*}=0.5$ and
various dipole moments: (left) $\mu^{*}=0$; (middle) $\mu^{*}=1$; (right)
$\mu^{*}=2$. In each case the symbols correspond to solute molecular
volumes of $V_{2}=V_{1}$ (circles), $V_{2}=2V_{1}$ (squares),
$V_{2}=3V_{1}$ (diamonds), and $V_{2}=4V_{1}$ (up triangles). The symbols
are from MC simulations and the curves are the predictions of
(\ref{eqn:mu2spt}).}
\end{figure}

\begin{figure}[!tbp]
\centering
\includegraphics[angle=270,scale=0.30]{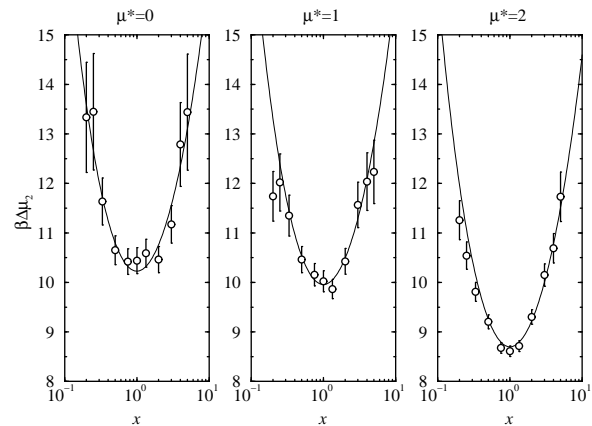}
\caption{\label{fig:r0.8vol} Excess chemical potentials of hard-spheroid
solutes with elongation $x$ and molecular volume $V_{2}=V_{1}$ in DHS
solvents with $\rho^{*}=0.8$ and various dipole moments: (left)
$\mu^{*}=0$; (middle) $\mu^{*}=1$; (right) $\mu^{*}=2$.  The symbols are
from MC simulations and the curves are the predictions of
(\ref{eqn:mu2spt}).}
\end{figure}

At higher densities very similar trends are observed in the results from
simulations and SPT. Results for $\rho^{*}=0.5$ and $\rho^{*}=0.8$ are
shown in Figs.~\ref{fig:r0.5vol} and \ref{fig:r0.8vol}, respectively. At
$\rho^{*}=0.5$ the simulation and SPT results clearly show that a prolate
solute has a lower excess chemical potential than an oblate solute with
reciprocal elongation and the same molecular volume. This trend is not
obvious in the simulation results at $\rho^{*}=0.8$ because we were only
able to carry out calculations for $V_{2}=V_{1}$ (although the SPT
predictions for larger solute volumes will of course show this effect).
Other trends apparent in Figs.~\ref{fig:r0.2vol}, \ref{fig:r0.5vol}, and
\ref{fig:r0.8vol} include the increase in $\beta\Delta\mu_{2}$ with
molecular volume and, more significantly, the decrease in
$\beta\Delta\mu_{2}$ with increasing $\mu^{*}$ or decreasing $T^{*}$.
Hence, the model system correctly shows that the solubilities of non-polar
solutes in polar solvents (or at least those that can be described
faithfully with a DHS model) will decrease with increasing temperature.
This variation implies that $\partial\ln{K}/\partial T^{*} < 0$, and hence
that the energy change $\Delta u_{2} < 0$ (\ref{eqn:deltau2}). Since
$\partial(T^{*}\ln{K})/\partial T^{*}<0$, it is also clear that the
entropy change $\Delta s_{2} < 0$ (\ref{eqn:deltas2}). The magnitudes of
the variations in $\beta\Delta\mu_{2}$ with temperature -- and hence
$|\Delta u_{2}|$ and $|\Delta s_{2}|$ -- increase with increasing solute
molecular volume. These results are analogous to those for hydrophobic
solvation, in which the presence of a solute molecule in the solvent
causes some sort of ordering of nearby solvent molecules with an
associated decrease of the solvent-solvent interaction energy. It should
be emphasized that in the present case, there are no solute-solvent
interactions beyond the hard-core repulsion, and so an energy change upon
inserting a solute molecule in to the DHS solvent would arise exclusively
from solvent-solvent interactions. In contrast to real hydrophobic
solvation, however, we anticipate that there will be no temperature
minimum in the solubility; the hard-core repulsions between model solutes
and solvents are never small compared to the thermal energy, and so
$\beta\Delta\mu_{2}$ will level off asymptotically to the values presented
here for the systems with $\mu^{*}=0$.

\section{Conclusions}
\label{sec:conclusions}

We have investigated the effects of molecular shape on the solvation of
hard-core solutes in model polar solvents. The solutes are hard spheroids
with a range of elongations and molecular volumes, while the solvent
consists of a fluid of dipolar hard spheres. Although these are very crude
representations, the properties of the model systems should bear some
qualitative resemblance to those of real solutions.

Using computer simulations we have calculated the distributions of
transient spheroidal cavities with specific elongations in the pure model
solvents. For a given solvent molecular dipole moment, the most probable
cavity dimensions decrease with increasing solvent density reflecting the
reduction in free volume. More interestingly, for oblate spheroidal
cavities the shape corresponding to the maximum average radial semi-axis
become less anisotropic with increasing solvent density; the same is true
for prolate spheroidal cavities with respect to the maximum average polar
semi-axis. This suggests that at high solvent densities, the probability
of developing a non-spherical cavity is reduced due to tight packing of
the solvent molecules. For a given solvent density and cavity elongation,
the average cavity size decreases with increasing solvent dipole moment
(or decreasing temperature) due to the association of solvent particles
and hence a reduction in the nearest-neighbor separation.

The excess chemical potentials of hard-spheroid solutes of various
molecular volumes have been calculated; these are related to the cavity
distributions since there is no solute-solvent interaction beyond the
hard-core repulsion. For given solute molecular volume, solvent density,
and temperature, the excess chemical potential is minimized for spherical
solutes, while this function is less for a prolate solute with elongation
$x$ than it is for an oblate solute with elongation $1/x$. These
observations are easily explained with reference to a simplified
scaled-particle theory written in terms of the equation of state of the
pure solvent, the solute-solvent interfacial tension, and fundamental
measures of the solute molecule, i.e., the mean radius of curvature, the
surface area, and the molecular volume. Firstly, the unfavorable
interfacial tension contribution is minimized for spherical solutes.  
Secondly, the violation of $x \leftrightarrow 1/x$ symmetry for a given
molecular volume is due to the differences in the mean radius of curvature
and surface areas between the two shapes. Since the solubility of a
spheroidal particle is simply related to the excess chemical potential, we
can predict that for a given molecular volume, rod-like particles should
be more soluble than plate-like particles, and spherical particles should
be the most soluble. To illustrate the magnitude of this effect, consider
a DHS solvent with $\rho^{*}=0.8$ and $\mu^{*}=1$, which roughly
corresponds to a typical polar molecular liquid. For prolate and oblate
solutes with moderate elongations of $3$ and $1/3$, respectively, and with
molecular volumes equal to that of a single solvent molecule,
Eqs.~(\ref{eqn:K}) and (\ref{eqn:mu2spt}) with the values given in Table
\ref{tab:thermo} predict that the solubility of the prolate solute will be
about 17\% higher than that of the oblate solute. Doubling the solute
volume would result in the prolate molecule being approximately 35\% more
soluble than the oblate one.

For given solute elongation and solvent density, the excess chemical
potentials of the solutes increase with increasing molecular volume, and
decrease with increasing solvent polarity (or decreasing temperature). The
former trend is obvious, while the latter trend is due to the
concentration of free volume mentioned above. The decrease in solubility
with increasing temperature is similar to that observed in the solvation
of hydrophobic solutes in water, and reflects the fact that the entropy of
solvation is negative, i.e., the presence of the solute molecule would
lead to an ordering of the solvent. As a result of this enhanced ordering,
the energy change is also negative. In the present calculations, the
negative energy change is entirely due to solvent-solvent interactions.

Unfortunately, a rigorous test of the simulation and theoretical results
against experimental data is not yet feasible due to the restricted set of
available molecular geometries and sizes; the key results presented here
concern hard solutes of equal volume and with reciprocal aspect ratios. It
would therefore be interesting to examine experimentally the solubilities
of, for example, nanoscopic colloidal particles with well controlled
geometries and particle volumes that are large compared to the solvent so
that the prolate-oblate asymmetry is enhanced.

\end{document}